\documentclass[proof]{WileyASNA-v1}
\articletype{Article Type}%

\received{26 April 2016}
\revised{6 June 2016}
\accepted{6 June 2016} 
\usepackage{natbib}
\usepackage[utf8]{inputenc}
\usepackage[T1]{fontenc}
\setcitestyle{square, comma, numbers,sort&compress, super}
\begin{document}

\title{Sub-second variability in black-hole X-ray binary jets}

\author[1,2,3]{Federico M. Vincentelli}

\author[3]{Piergiorgio Casella}

\authormark{Vincentelli \& Casella}

\address[1]{Department of Physics \& Astronomy, University of Southampton, Highfield, Southampton SO17 1BJ, UK}

\address[2]{INAF, Osservatorio Astronomico di Brera Merate, via E. Bianchi 46, I-23807Merate, Italy}

\address[3]{INAF, Osservatorio Astronomico di Roma, Via Frascati 33, I-00078 Monteporzio Catone, Italy}

\corres{\email{federico.vincentelli@inaf.it}}


\abstract{In the last 10 years multi-wavelength fast variability studies of low mass X-ray binaries  have shown a dramatic development. A key discovery was the detection of O-IR sub-second fluctuations in two black-hole transients, lagging the X-rays by $\approx$0.1 s. This demonstrated how the fluctuation observed in the inflow could be transferred to the jet, allowing therefore also to study in a completely new way the physical processes which take place at the base of the jet. In this paper we review the latest developments of the study of jets with this new approach, focusing on the results obtained with cross-spectral analysis techniques.}
\keywords{black-hole, jets, accretion, X-ray binaries}



\maketitle


\section{Introduction}\label{sec1}

Black-hole low mass X-ray binaries (BH LMXRBs), are transient sources which are known to present emission along almost all the electromagnetic spectrum, from radio to hard X-rays. Systematic studies in X-rays showed that during their outburst these sources can display two main spectral state: a soft and a hard state. In the first case, the X-ray spectrum can be described by the black-body emission from an optically thick, geometrically thin accretion disk: in the second, the spectrum is dominated by a power-law up to $\approx$ 100 keV which has been explained in terms of Comptonization of the disk photons from a hot optically thin, geometrically thick inflow \citep{dgk07}.   It has also been shown that during this state, BH LMXRBs display radio emission with an approximately flat spectrum which can extend up to optical-infrared (O-IR) frequencies \citep{corbel2002,corbel2013,russell_d2013}.  Such feature has been successfully modelled with the superposition of the syncrothron emission profiles generated by the acceleration of the electrons in a relativistic jet \citep{blandford&konigl}. In order  to reproduce the observed spectral energy distribution, such model assumes that the energy of the electrons is continuously replenished, so that the radiative and adiabatic energy losses are balanced.  However, the origin of the process which provides such energy is still a matter of debate.

To solve this issue, in analogy to what is generally believed to happen in Gamma-Ray Bursts \citep{kobayashi1997} and in blazars \citep{spada2001}, different models tried to apply the internal shocks scenario also to the jets in these sources. According to such scenario, shells of matter are launched into the jet with variable velocities. Due to differences in velocity they collide and merge. Application to XRBs showed that through the conversion of kinetic to internal energy in the shocks, this process can balance the energy losses and reproduce the observed flat spectrum at lower energies\citep{jamil2010}. Of course, the key prediction of such scenario is the presence of variable emission from the jet.


First evidences of strong variability from BHXRB jets were found in GRS 1915+105.  Multiwavelength observations revealed the presence of spectacular infrared and radio flares on timescales of $\approx$ hours \citep{mirabel1998,eikenberry1998,fender1998,fender2000,rothstein2005,lasso-cabrera2013}. These variations, however,  were found to be due to relativistic ejections launched from the system, in correspondence to strong accretion episodes, and not to the presence of shocks in a compact jet. Nevertheless the observation of this behaviour increased the scientific attention towards the variability at lower energies, initiating the modern approach to this kind of studies. 

In the early 2000s, new fast read-out detectors allowed to probe with sub-second time resolution the optical emission of these sources. The first simultaneous sub-second optical/X-ray observations revealed a complex connection between the emission in the two bands \citep{kanbach2001,hynes2003,hynes2006,durant2008}. The optical response to the X-ray fluctuations could not be explained just with the emission from a compact jet \citep{malzac2004}.  Such complexity derives from the presence of different processes which can take place in the optical band (e.g. synchrotron radiation from the hot inflow \citep{veledina2011,veledina2013}, or reprocessing from the outer disk \citep{obrien}).

A crucial step in the study of jets is represented by the detection of a 0.1 s lag optical-infrared (O-IR) emission in respect to the X-rays during the hard state of GX 339-4. Such measurement showed how the fluctuations from the inflow can be transferred into the compact jet in fractions of a second \citep{casella2010,gandhi2008,kalamkar2016}.  This was then confirmed with the recent detection of a similar lag in V404 Cygni, which also allowed to put new constrains on the size of the jet base \citep{gandhi2017}. Moreover such discovery opened the possibility to study in detail the physical processes which take place inside the jet, through the analysis of the fast multiwavelength variability \citep{gandhi2010,malzac2018,vincentelli2018}. The aim of this paper is to review the latest developments in the study of the jet fast variability. Therefore, even though fast optical observations have been key to define the phenomenological scenario, we choose to focus only on the results found in the IR band, which are less affected by other components and allowed to reach a deeper insight on the jet's internal processes. 

\section{Fast IR variability}\label{sec2}

\subsection{First results}

The first unambiguous detection of fast variability from a compact jet was found in the IR in GX 339-4 during the so called low-hard state  \citep{casella2010}. The fast infrared observation presented variability down to $\approx 200$ ms timescale, which, with brightness temperature arguments, excluded any possibility of emission from the inflow. Moreover also the measured 0.1 s lag in the cross-correlation function (CCF), presented a symmetric structure, inconsistent with being reprocessing from the outer disk. 

Such result, together with the discovery in the same years of a similar lag also between the optical and the X-ray variability in GX 339-4 \citep{gandhi2008,gandhi2010}, had great relevance from the theoretical point of view. Evidence of fast variability from the base of the jet strongly suggested that internal shocks could effectively play a crucial role also in BH XRBs.  This was confirmed by the model developed by \cite{malzac2013,malzac2014}: by linking the velocities of the injected shells to the fluctuations measured in the X-ray luminosity, such model managed to reproduce not only the sub-second O-IR variability, but also the observed 0.1 s lag.

Further observations were then performed to investigate how the inflow/outflow connection could evolve during the outburst. In particular, this was done by \cite{kalamkar2016} who observed the same source during a higher luminosity state, near to the transition to the soft state (also known as high-intermediate state). This brought to the discovery of the first IR Quasi-periodic oscillation (QPO), which resulted also in harmonic relation with a simultaneous QPO in the X-rays. The CCF together with the usual 0.1 s lag, showed also evidence of some oscillations.  While the fast variability associated to the 0.1 s lag was easily associated to fluctuations in the jet, the QPO could not be explained in terms only of a hot inflow precessing, and therefore was interpreted as the effect of a jet precessing together with the hot inflow.

\subsection{Deeper analysis}

Given the physical interpretation of these results, the described observations could also be used to investigate how the jet modifies the fluctuations which are injected from the inflow. In particular, this was done by \cite{vincentelli2018} and \cite{malzac2018} by applying Fourier domain  techniques (i.e.  cross-spectral analysis) to simultaneous X-ray/IR light curves. The two main quantities which can be measured are the the coherence and phase/time lags (For a review see \cite{uttley2014}). While the first gives a measure of the degree of linear correlation between two curves, the seconds estimate a possible lag as a function of the Fourier frequency.   Fig. 1 shows these quantities for the two datasets analysed in GX 339-4.


\begin{figure*}
\centering
	\centerline{\includegraphics[width=\textwidth]{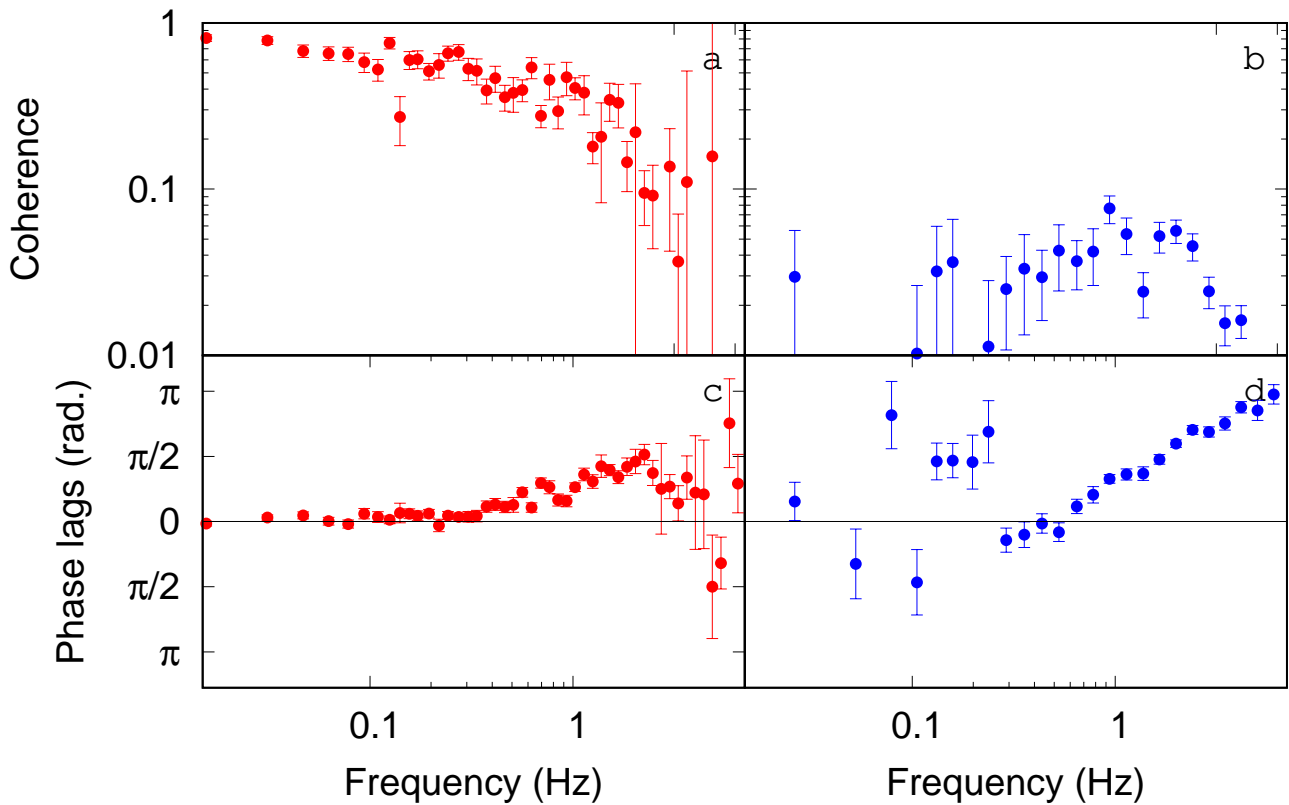}}
	\caption{Comparison between the cross-spectral properties during the low-hard state  (red) and high-intermediate state (blue) (From\cite{vincentelli2018} and  \cite{malzac2018} respectively). Main difference can be seen at low frequencies: while in the first case the coherence is high and decreases gradually (a), in the second, it significantly lower at all frequencies (b). Moreover phase lags of the high-intermediate state (d) present a negative value of $\approx$-$\pi/2$, possibly "polluted" by the presence of the QPOs.  }
\label{figure}
\end{figure*}
 
As expected from the shape of the CCF, in the low-hard state a simple constant time lag at 0.1 s was measured\footnotetext{A constant time lags becomes a linearly increasing phase lag.} \citep{vincentelli2018}. The coherence, instead, showed a rather particular trend, decreasing gradually as a function of frequency.  Such result indicated that the way fluctuations are transferred into the jet cannot be seen as a simple linear impulse response function (see \cite{nowak1999} and \cite{uttley2014}), suggesting the presence of some kind of non-linearity as the origin of the IR emission. However, given that a classical non-linear process would just give a very low coherence, the authors suggested that the smooth decrease of the coherence could be due to a non-stationarity of the system, which leads to a broad IR-emitting region in the jet. Such scenario would also be in agreement with the internal shock models, as the region where shock takes place, and therefore variability is emitted, is not stably localized (see also Fig. 4 in \citep{malzac2018}).
 
The cross-spectral analysis measurements done in the higher luminosity state (performed by \cite{malzac2018}) were found to be significantly different. In particular, the lags showed a complex and non-intuitive trend. While at high frequency the "usual" 0.1 s lag was clearly detected, at lower frequency the lags changed sign tending at $\approx$ -$\pi/2$.  Moreover, a distinct component was also seen in correspondence of the two QPOs. Such complex lags were found to be associated to very low coherence, indicating the presence of a highly non-linear process. Despite the strong differences, the authors found that such peculiar trend could be still naturally explained by internal shocks:  the origin of the negative lag was identified as a consequence of the Doppler boosting modulation. Such relativistic effect leads to an anti-correlation between X-ray and IR on longer timescales, generating a constant negative phase lag. Such effect is partially distorted, due the presence of two QPOs at 0.08 and 0.16 Hz, but still visible.  

 \section{Discussion}\label{sec1}

Recent application of cross-spectral analysis to simultaneous sub-second IR/X-ray observations of GX 339-4  permitted to shed light on the physical process which take place inside the jet.  In particular, the observations seem to be in good agreement with the expectations from internal-shock models.  Moreover thanks to the application of Fourier techniques it has been possible for the first time to appreciate in the IR variability effects due to the relativistic motion of matter accelerated in the jet.

The significant differences observed in the two states can also give indications on the behaviour of the jet.  For example the appearance at higher luminosities  of an anti-correlation on longer timescales, suggests some changes in the physical/geometrical conditions of the jet, which enhance the Doppler boosting effect.  Given that the inclination of the source cannot change, a possible explanation for this behaviour could be the change in the average $\Gamma$ Lorentz factor which leads to a smaller viewing angle of the jet itself. This would then introduce an anti-correlation as the jet points away from our line of sight. It is interesting to notice that a recent application of this same model to multi-epoch observations of  MAXI J1836-194 (Peault et al. in prep.) found similar results, requiring  an increase of the average jet velocity during the evolution of the outburst.

 \section{Conclusions}\label{sec2}

We gave a synthetic review of the latest results on fast variability from the jet in BH LMXRBs. While the first fast IR observations managed mainly to put strong constrains on the geometry of the jet, recent application of cross-spectral analysis techniques are permitting to probe the internal processes of the jet.  Two observations performed at different stages of the outburst suggest that the jet is evolving together with the inflow properties.  The main differences seem to be related to relativistic effects, which would then indicate an increase in the jet average velocity. Multiple observations along the same outburst are needed to  put much stronger constrains on the physical scenario. The great effort put in the coordination between observing facilities is now facilitating the building of multi-wavelength campaigns of transient sources. New campaigns are indeed now managing to follow with several facilities the evolution of the multi-wavelength variability of these source, and will give in the near future a further significant contribution to the understanding of jets.

\bibliography{bib.bib}{} 
\end{document}